# Understanding the difference in cohesive energies between alpha and beta tin in DFT calculations

Fleur Legrain[a] and Sergei Manzhos[a]

[a]Department of Mechanical Engineering, National University of Singapore, Block EA #07-08, 9 Engineering Drive 1, Singapore 117576, Singapore. E-mail: mpemanzh@nus.edu.sg (SM).

The transition temperature between the low-temperature alpha phase of tin to beta tin is close to the room temperature ($T_{\alpha\beta}$ =13$^0$C), and the difference in cohesive energy of the two phases at 0 K of about $\Delta E_{coh}$=0.02 eV/atom is at the limit of the accuracy of DFT (density functional theory) with available exchange-correlation functionals. It is however critically important to model the relative phase energies correctly for any reasonable description of phenomena and technologies involving these phases, for example, the performance of tin electrodes in electrochemical batteries. Here, we show that several commonly used and converged DFT setups using the most practical and widely used PBE functional result in $\Delta E_{coh}$≈0.04 eV/atom, with different types of basis sets and with different models of core electrons (all-electron or pseudopotentials of different types), which leads to a significant overestimation of $T_{\alpha\beta}$. We show that this is due to the errors in relative positions of *s* and *p* –like bands, which, combined with different populations of these bands in α and β Sn, leads to overstabilization of alpha tin. We show that this error can be effectively corrected by applying a Hubbard +*U* correction to *s* –like states, whereby correct cohesive energies of both α and β Sn can be obtained with the same computational scheme. We quantify for the first time the effects of anharmonicity on $\Delta E_{coh}$ and find that it is negligible.

## Introduction

Tin (Z=50) is part of the group IV materials. Similarly to its lighter counterparts (C, Si, and Ge), Sn exists in the diamond-type crystal structure, commonly called alpha Sn (α-Sn, space group 227; Fd-3m) or grey tin; this is the most stable low-temperature phase, and it is a zero-gap semiconductor.[1] Unlike C, Si, and Ge, however, the diamond structure of Sn is only stable under 286 K (13°C), above which Sn transforms to beta Sn (β-Sn, space group 141; I41/amd), the metallic and room temperature phase of tin. The relatively low transition temperature $T_{\alpha\beta}$ is reflective of an only slight stabilization of α Sn over β Sn at 0 K (estimated at about $\Delta E_{coh}$=0.02 eV/atom), so that relatively small vibrational contributions are required to reverse the phase stability. The proximity of $T_{\alpha\beta}$ to room temperature means that the α–β transition can be of practical importance, as the relative phase stability can be easily changed by perturbations to the lattice by e.g. doping or strain.

An example of a technology where this transition has been found important are electrochemical batteries. Tin has been shown to be a high capacity anode material for Li, Na, and Mg ion batteries.[2-5] Specifically in Li ion batteries, for which Sn-based electrodes have been most studied,[6,7] the formation of α Sn upon lithiation of (initially β phase) Sn electrodes has been reported in experimental studies.[8,9] This beta-to-alpha phase transformation in battery electrodes is still not fully understood. To produce reliable *ab initio* models of this process or other processes which can affect the phase stability at room temperature, the transition temperature and therefore $\Delta E_{coh}$ need to be reproduced accurately.

The practical approach to *ab initio* modelling of solids is DFT (density functional theory) with PBC (periodic boundary conditions), and GGA (generalized gradient approximation)[10] with the PBE (Perdew–Burke–Ernzerhof)[11] functional is the most practical and widely used approximation to the exchange-correlation functional.[12,13] Hybrid functionals[14,15] can be used with small simulation cells[16] but are increasingly unwieldy for larger supercells which are required to study doping, intercalation, and interfaces. A difference in cohesive energy on the order of 10$^{-2}$ eV/atom is at the limit of DFT accuracy with available exchange-correlation functionals. Previous DFT simulations reported $\Delta E_{coh}$ of 0.022-0.047 eV/atom.[17,18] Below, we show that *converged* calculations using the PBE functional and different types of basis sets and core electron treatment invariably result in $\Delta E_{coh}$≈0.04 eV/atom. This leads to an overestimation of $T_{\alpha\beta}$ by more than 100 K if the harmonic approximation for the phononic contribution is used and is therefore not suited for modelling near-room temperature behaviour of tin-based materials.

The transition temperature is the temperature at which the sum of the electronic ($E_{DFT}$) and of the vibrational energy ($E_{vib}$) and entropy ($-TS$) contributions,

$$G = E_{DFT} + (E_{vib} - TS), \qquad (1)$$

is equal in the two phases. Eq. (1) is an approximation for the Gibbs free energy that neglects the *pV* term; this is appropriate for low pressure conditions including standard conditions. At the transition temperature, the difference in vibrational contributions to $G$ counterbalances exactly the DFT energy difference, i.e. $\Delta E_{coh} = \Delta(E_{vib} - TS)$. This means that the error in $T_{\alpha\beta}$ can come either from the error in electronic energy or/and in the vibrational contribution, on the order of 0.02 eV/atom. The difference in vibrational contribution to $G$ at the experimental transition temperature between alpha and beta Sn is around 0.024 eV based on GGA DFT and the harmonic approximation (see below).

With an accurate computational setup, each of the two terms of Eq. (1) should therefore be exact to within a few meV, and not within a few dozen meV as is the case with the standard setup using the PBE functional and the harmonic approximation. It is also clear that an accurate setup must be able to model correctly *both* α and β Sn simultaneously to model systems where the phase transition is possible. It means that fixes such as selecting the degree of exact exchange[15,19] or Hubbard (+$U$) corrections[20] that tune the cohesive energy and band structure with parameters different for the two phases are not useful.

In this paper, we investigate the origin of the error in $T_{\alpha\beta}$ and identify a computational scheme that fixes it. We analyse both the electronic and the phononic contributions to free energy. Specifically, for the first time, we analyse the effect of anharmonicity of vibrations of bulk tin on relative phase stability; it is shown to be unimportant. We show that the computed $\Delta E_{coh}$ of about 0.04 eV is ubiquitous across DFT setups using localized or plane wave based basis sets and full ionic potential or pseudopotentials. We trace the origin of the error in the electronic energy to destabilization of the energy of the *s* band vs the *p* band and show that the correct $\Delta E_{coh}$ can be obtained by applying a Hubbard +$U$ correction to the atomic-part *s* orbitals, with which absolute and relative cohesive energies of both alpha and beta tin can be accurately modelled with one and the same computational setup.

## Methods

DFT calculations were performed with the Perdew-Burke-Ernzerhof (PBE) exchange-correlation functional.[11] Different codes were used employing (augmented) plane wave (VASP[21] and Elk[22]) or atom-centered (SIESTA[23] and FHI-AIMS[24]) basis sets, full ionic potential (FHI-AIMS and Elk) or pseudopotentials (VASP and SIESTA). When using pseudopotentials, the valence configuration was $5s^25p^2$; the effect of the addition of the 4d electrons to the valence shell was found to be insignificant. Spin polarized calculations were used where spin polarization is important (i.e. tin atoms and dimers) and spin restricted where it is not (i.e. bulk phases).

The VASP[21] calculations were performed using a PAW (projector-augmented wave) pseudopotential[25] and a plane wave energy cutoff of 300 eV. Cubic 8 atom cells were used for both α and β Sn. 16×16×16 Γ-centred Monkhorst-Pack meshes were used for the *k*-point sampling of the Brillouin zone integration for α and β Sn. A Gaussian smearing is used with σ=0.005. Atomic coordinates and cell vectors were optimized until the residual forces were below 0.01 eV/Å. To consider the vibrational contributions, we computed the Hessian matrices on cubic 64-atom cells (8×8×8 k-mesh) using the displacement method and a step of 0.01 Å.

The Elk[22] calculations were performed using the full potential linearized augmented plane wave method (FP-LAPW). This approach implements a solution to an all-electron and a full-potential system. The 36 energetically lowest electrons are considered to be core electrons and the remaining 14 electrons are valence electrons. The core electrons are treated fully relativistically in the spin-compensated Dirac equation for spherical potential, and the valence electrons are treated within the scalar-relativistic approximation where the spin-orbit coupling is included as a perturbation. The muffin-tin radius is $R$ = 2.3 a.u.[26] The plane-wave cutoff $K_{max}$ was determined from $R \cdot K_{max}$ = 10.0 a.u. The angular momentum cutoff for the APW functions and for the muffin density and potential was set to 10 and 9, respectively. Maximum length of the reciprocal vectors $G_{max}$ for the expansion of the interstitial density and potential was get to 14.0 a.u. Total energy convergence was set to <0.5×10$^{-6}$ a.u./atom, and the electron population function was broadened with the Fermi-Dirac distribution with width 0.001 a.u. Eight- and four-atom cells were used for α and β Sn, respectively, and 10x10x10 and 10x10x20 *k* points, respectively, provided converged results.

The SIESTA[23] calculations were performed using a Troullier-Martins pseudopotential[27] (generated by us) and DZP and TZP (double- ζ polarized and triple-ζ polarized, respectively) basis sets. The DZP basis was tuned by using the *PAO.EnergyShift* parameter to reproduce both the absolute value of $E_{coh}$ (to the experimental value, i.e. ≈3.14 eV/atom)[28] and the difference between the phases, $\Delta E_{coh}$ (to the converged plane wave values, i.e. 0.04 eV/atom). We then checked that the conclusions hold for a TZP basis made with the same value of *PAO.EnergyShift* and are therefore not polluted by a particular choice of the atom-centred basis. A 200 Ry cutoff was used for the expansion of the density with a *bcc*-type oversampling of the Fourier grid. Eight- and four-atom cells were used for α and β Sn, respectively, with 10x10x10 and 10x10x20 *k* points, respectively, for most calculations; we confirmed that doubling the number of *k* points in each direction does not change the result. Atomic coordinates and cell vectors were optimized until the forces on atoms were below 0.01 eV/Å and stresses below 0.1 GPa. A development version of SIESTA which includes the LDA+U functionality[29] with the Dudarev approximation[30] was used to apply Hubbard correction on *s* or *p*–like levels.

The FHI-AIMS[24] calculations were performed with basis sets and integration grids set to the "really_tight" settings to approach the basis set limit. The convergence criteria for the self-consistency cycle were 1×10$^{-7}$ eV for the energy and 1×10$^{-6}$ for the density. Unit cells were used (with 2 atoms per cell, i.e. non-rectilinear). A smearing of σ=0.1 eV was used. 15 *k*-points per dimension provided converged Brillouin zone sampling (change in energy of either phase of less than 0.001 eV vs. 30 *k*-points). Atomic coordinates and cell vectors were optimized until the residual forces were below 1×10$^{-2}$ eV/Å. The atomic ZORA[31] approximation was used to account for relativistic effects.

CCSD(T)[32] calculations were performed in Gaussian 09 (G09).[33] Calculations were performed on Sn atoms and dimers using the LANL2DZ[34,35] basis set, and the trends in the differences in $s$ and $p$ levels between CCSD(T) and DFT were also confirmed with calculations on atoms using a larger QZVPPD basis set with a (small-core) ECP28MDF ECP.[36] DFT calculations were also performed in Gaussian 09 on atoms and dimers for comparison with CCSD(T) calculations.

Contributions to the free arising from the vibrational energy and entropy were computed as:

$$E_{vib} - TS = \sum_{i=1}^{N}\left\{\tfrac{1}{2}v_i + k_B T \ln(1 - e^{-v_i/k_B T})\right\} \quad (2)$$

where $v_i$ is the energy of one quantum in the $i$-th normal mode, $T$ is the temperature, and $k_B$ is the Boltzmann constant. The sum is over all $N$ normal modes. Adding these vibrational contributions to the DFT ground state energy $E_{DFT}$ gives the free energy of the system, Eq. (1).

Molecular dynamics (MD) simulations were performed in VASP with a time step of 2 fs. Cubic cells with 64 Sn atoms were used for both phases. The $k$-grid was reduced to 3x3x3 for alpha and 4x4x4 for beta Sn. The temperature was ramped from 0 K to 300 K over 2000 (3000) steps for β (α) Sn. A total of 15,000 steps were computed for α and 10,000 steps for β Sn.

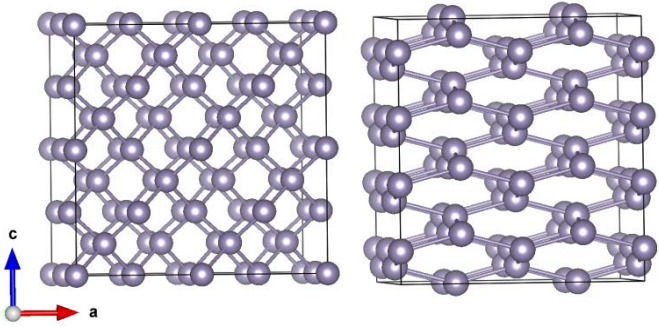

**Fig. 1** Crystal structures of alpha (left) and beta tin.

## Results

### Alpha and beta tin with different DFT setups

Crystal structures of α and β tin are shown in Fig. 1. Table 1 lists α Sn cohesive energies as well as the differences in cohesive energies between α and β Sn obtained with different DFT codes, basis set types and core electron treatments. A difference in cohesive energies of about 0.04 eV/atom is persistent among different computational setups and is different by about 0.02 eV/atom from experimental estimates of about 0.02 eV/atom.[37] This results in a significant overestimation of the transition temperature, e.g. value $T_{\alpha\beta}$ = 475 K is obtained in VASP using the harmonic approximation, Eq. (2). We have confirmed that using the quasiharmonic approximation, namely, fixing the cell volume according to the thermal expansion coefficient of Sn, does not significantly affect the result, changing the transition temperature by only 45 K (as computed in VASP). The error in $\Delta E_{coh}$ and $T_{\alpha\beta}$ could come mainly from the electronic or vibrational contributions to the free energy, or both. In what follows, we therefore analyse each of these contributions.

**Table 1**. Cohesive energies of α Sn and the difference in $E_{coh}$ (eV) between α and β Sn, $\Delta E_{coh}$ (eV) as well as lattice parameters ($a$ for α Sn and $a, c$ for β Sn, in Å) with different computational schemes (all using the PBE functional). Units of $U$ are eV.

| Setup | $E_{coh}$ (α) | $\Delta E_{coh}$ | $a$ (α) | $a, c$ (β) |
|---|---|---|---|---|
| VASP | 3.20 | 0.039 | 6.65 | 5.93, 3.22 |
| SIESTA[a] | 3.17 | 0.044 | 6.68 | 5.95, 3.25 |
| SIESTA U(s)=+1.0[a] | 3.12 | 0.026 | 6.69 | 5.96, 3.26 |
| SIESTA U(s)=+1.5[b] | 3.13 | 0.025 | 6.71 | 5.98, 3.26 |
| SIESTA U(s)=+1.0[c] | 3.17 | 0.022 | 6.71 | 5.98, 3.23 |
| SIESTA U(s)=+1.25[c] | 3.15 | 0.018 | 6.71 | 5.99, 3.22 |
| VASP U(s)=+1.0 | 3.16 | 0.023 | 6.66 | 5.94, 3.22 |
| VASP U(s)=+1.5 | 3.13 | 0.015 | 6.67 | 5.95, 3.22 |
| Elk |  | 0.045 | 6.65 | 5.98, 3.20 |
| FHI-AIMS | 3.18 | 0.040 | 6.65 | 5.94, 3.20 |
| experiment | 3.14[d] | 0.02[e] | 6.49[e] | 5.82[e]-5.83[f], 3.18[f] |

[a] PAO.EnergyShift=0.001 Ry, DZP basis
[b] PAO.EnergyShift=0.002 Ry, DZP basis
[c] PAO.EnergyShift=0.002 Ry, TZP basis
[d] Ref. 28. The cohesive energy difference is corrected for the ZPE.
[e] Ref. 37.
[f] Ref. 38. P.310.

### Effects due to anharmonicity

The vibrational part of free energy as expressed in Eq. (2) neglects the effects of anharmonicity and coupling on vibrational levels. To estimate these effects, we performed MD simulations of both phases of tin to sample the potential energy surface up to potential energies relevant for $T_{\alpha\beta}$. The collected potential values, $V_{DFT}$, were compared to the harmonic uncoupled potential built on normal mode frequencies:

$$V_{harm} = \sum_{i=1}^{N} \tfrac{1}{2} v_i^2 Q_i^2 \quad (3)$$

where $Q_i$ are normal mode coordinates related to the Cartesian coordinates of Sn atoms via

$$\mathbf{Q} = \mathbf{L}^{-1} \mathbf{M}^{\tfrac{1}{2}}(\mathbf{x} - \mathbf{x}_{eq}) \quad (4)$$

where $\mathbf{Q}$ is a vector of all normal mode coordinates, $\mathbf{L}$ is the matrix that diagonalizes the Hessian matrix, $\mathbf{x}$ is a vector of Cartesian coordinates of all atoms, and $\mathbf{x}_{eq}$ are equilibrium positions. $\mathbf{M}$ is a diagonal matrix of atomic masses. We consider $N=3N_A$, where $N_A$ is the number of atoms in the simulation cell, in which case all matrices are square of size $N \times N$ and vectors are of length $N$. This therefore includes (zero-frequency) frustrated translation modes.

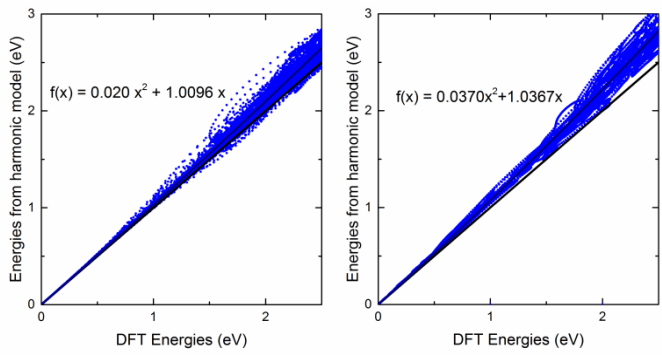

**Fig. 2** Harmonic (eq. 3) vs. anharmonic (DFT values along the MD trajectories) potential energies following from the MD simulations for alpha (left) and beta tin.

Fig. 2 shows the plots the harmonic potential energies $V_{harm}$ vs. anharmonic energies ($V_{DFT}$) following from the MD simulations for both phases. The onset of anharmonicity around room temperature (corresponding to $V \approx 2.5$ eV) can be visually appreciated. Does it significantly contribute to relative phase stability? In the classical limit (i.e. all $v_i \to 0$) there would be no effect of anharmonicity and coupling in the potential on relative phase stability, while in the limit of large $v_i$ (specifically, when $v_i \gg k_B T$), changes in $v_i$ would directly contribute to changes in $G$. The spectrum of $v_i$ spans the range up to about 175 cm$^{-1}$ for $\alpha$ and up to about 130 cm$^{-1}$ for $\beta$ Sn i.e. all components are smaller than or on the order of $k_B T$ at room temperature (see Fig. 3 for phonon DOS).

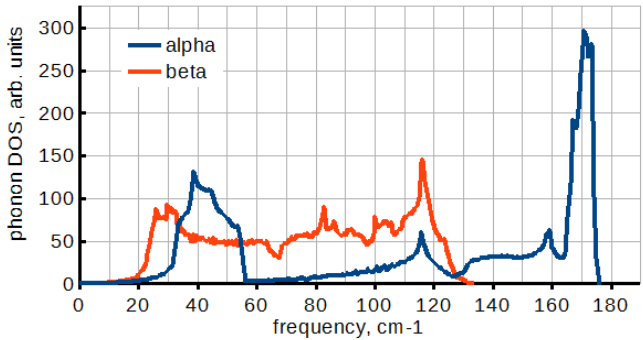

**Fig. 3** Phonon density of states for $\alpha$ and $\beta$ Sn.

To estimate the effect of anharmonicity on $G$, we express Eq. (2) as

$$E_{vib} - TS = -k_B T \sum_{i=1}^{N} \ln\left(\sum_{n_i=0}^{n_{i,max}} \exp\left(-\frac{\epsilon_{i,n_i}}{k_B T}\right)\right) \quad (5)$$

where the inner summation uses a sufficiently large $n_{i,max}$ for convergence, and $\epsilon_{i,n_i}$ are the vibrational energies with $n_i$ quanta of excitation in mode $i$. We perform a non-linear regression of the dependence of $V_{harm}$ on $V_{DFT}$, $V_{harm} = f(V_{DFT})$ as shown in Fig. 2 and apply the function $f$ to estimate $\epsilon_{i,n_i}$:

$$\epsilon_{i,n_i} = f^{-1}\left(v_i\left(n_i + \frac{1}{2}\right)\right) \quad (6)$$

As changes in vibrational levels are typically only a fraction of changes in the potential energy values,[39] Eq. (6) can be considered as an upper limit of the effect due to anharmonicity. It results in a change in the relative energies of the two phases on the order of 0.001 eV/atom at room temperature. This would only change the transition temperature by about 10 K and is insignificant compared to the difference of about 0.02 eV/atom between the computed $\Delta E_{coh}$ and $\Delta(E_{vib} - TS)^{290K}$.

We therefore conclude that anharmonicity and coupling of tin vibrations are not the cause of the overestimation of $T_{\alpha\beta}$ reported in the previous subsection and that $\Delta E_{coh}$ must be overestimated by DFT.

**Analysis of electronic structure**

To understand why there is persistent overestimation of $\Delta E_{coh}$ with GGA DFT (PBE functional) regardless of the type of basis set or pseudopotential, we turn to the differences in electronic structure between alpha and beta Sn and analyse how well they are modelled by DFT. From the Mulliken population analysis (performed with atom-centered basis set calculations in SIESTA), we find that $\alpha$ Sn has a higher degree of transfer of electrons from $s$ to $p$ population than $\beta$ Sn when forming bonds in bulk tin: the $s$ population in $\alpha$ Sn is about 1.5$e$ and in $\beta$ Sn about 1.7$e$ (to compare to 2.0$e$ in a free atom, as per the valence shell configuration 5s$^2$5p$^2$). The differences in $s$ vs $p$ population directly contribute to the total energy, and thereby to $\Delta E_{coh}$, via the bandstructure component of the total energy. The population of orbitals with $d$ character upon bulk formation is found to be insignificant. Fig. 4 shows projected density of states for both phases of tin.

In Table 2, we compare the average energies of $s$ and $p$ bands (computed as population-weighted energies of states with $s$ and $p$ character) and the difference between them ($p$-$s$ "gap") in a Sn atom and in Sn$_2$ dimers computed with DFT to those computed with CCSD(T) as a reference. There is a similar pattern of differences (errors) in $s$ and $p$ –like band energies computed with DFT in all codes we used vs. CCSD(T); namely, both $s$ and $p$ –like levels computed with DFT are higher in energy (as expected) than the corresponding CCSD(T) values. More importantly, the $s$ –like level is destabilized by a larger amount, and therefore the $p$-$s$ gap is lower than with CCSD(T). When bonds are formed between tin atoms, in CCSD(T), the $s$ band energy is lowered by 0.83 eV while the $p$ band energy is slightly raised by 0.1 eV. In the DFT calculations, both $s$ and $p$ levels are stabilized by bonding, and $p$ levels are stabilized more than $s$ levels.

This qualitative difference is observed in the DFT calculations in all Gaussian 09, VASP, and SIESTA and is therefore not an artefact caused by a specific computational setup. In the periodic codes, we have confirmed that the $s$ and $p$ level energies are converged to within 0.01 eV in SIESTA and 0.025 eV in VASP with respect to the size of the cubic supercell (total energies and $s$-$p$ gaps are converged much faster with respect to cell size than eigenvalues). It is therefore expected that in materials where Sn-Sn bonding is important, $s$ and $p$ band energies will be in error, and by a different amount.

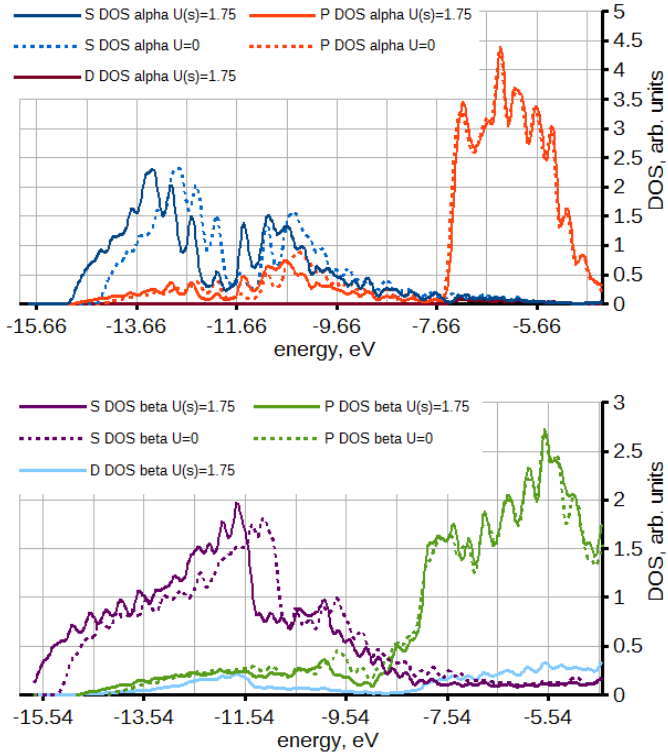

**Fig. 4** The density of states of alpha (top) and beta (bottom) Sn projected on $s$, $p$, and $d$ orbitals. The plot is up to the Fermi level.

If the stronger destabilization of $s$ vs $p$ band with DFT using the PBE functional which is observed in a Sn atom and Sn dimer carries over into tin bulk, then one expects that a phase with a higher $p$ population will be additionally (and erroneously) stabilized vs. a phase with a lower $p$ population. This is exactly what is observed in alpha vs. beta Sn.

Conversely, it must be possible to reverse this effect by stabilizing the $s$-like band. This can be achieved by applying a positive Hubbard correction or $U$ value to $s$-like states. The effect of lowering the $s$ band energy and of increasing the $p$-$s$ gap was indeed confirmed in Sn atom and dimer calculations. For example, applying $U(s)$ = 1.5 eV in SIESTA reduces the destabilization of the $s$ band of a Sn atom vs. CCSD(T) (Table 2) from 3.3 eV to 2.8 eV while the destabilization of the $p$ states remains at 3.1 eV; the $s$ band destabilization in the dimer is reduced from 4.1 eV to 3.6 eV and the degree of underestimation of the $p$-$s$ gap from 1.6 eV to 1.1 eV, while the $p$ band destabilization remained at 2.5 eV. A similar effect is observed in VASP (see Appendix). We will see below that a similar order of magnitude of $U$ is required to correct for the over-stabilization of the alpha vs. beta phase.

**Table 2**. Energies (in eV) of $s$ and $p$ –like bands (vs. vacuum) and their differences in a tin atom and tin dimer computed with different methods.

| band | CCSD(T) | | | DFT (G09) | | | DFT-CCSD(T) (G09) | | DFT(SIESTA[a])-CCSD(T) | | DFT(VASP)-CCSD(T) | |
|---|---|---|---|---|---|---|---|---|---|---|---|---|
| | Sn | $Sn_2$ | $Sn_2$-Sn | Sn | $Sn_2$ | $Sn_2$-Sn | Sn | $Sn_2$ | Sn | $Sn_2$ | Sn | $Sn_2$ |
| $s$ | -13.48 | -14.31 | -0.83 | -9.98 | -10.18 | -0.20 | 3.50 | 4.12 | 3.32 | 4.07 | 3.14 | 3.97 |
| $p$ | -7.08 | -6.98 | 0.10 | -4.16 | -4.69 | -0.53 | 2.93 | 2.30 | 3.07 | 2.46 | 2.97 | 2.40 |
| p-s | 6.40 | 7.32 | 0.93 | 5.82 | 5.50 | -0.35 | -0.57 | -1.83 | -0.26 | -1.61 | -0.17 | -1.57 |

[a] TZP basis, PAO.EnergyShift = 0.002 Ry

We have then applied different +$U$ values to $s$ –like states of both alpha and beta tin to test if this conjecture holds for bulk Sn. With the increase of the $U$ value, $\Delta E_{coh}$ indeed decreased. When using a localized basis set in SIESTA (which was tuned to $E_{coh}$ of α Sn), the application of +$U$ led to a modest weakening of $E_{coh}$ (within 0.1 eV/atoms for $U$ values of up to +2.0 eV) and only a small change of the lattice constant (on the order of 0.01 Å), see Table 1. This can easily be compensated by tuning the PAO.EnergyShift parameter. For example, with a DZP basis obtained with PAO.EnergyShift=0.002 Ry and giving $E_{coh}(\alpha)$=3.13 eV/atom (with +$U$), a $U$ value of +1.5 eV results in $\Delta E_{coh}$ = 0.025 eV/atom. To confirm that the effect of +$U$ is not an artefact of the choice of the atom-centered basis set, we performed a calculation with a TZP basis (using the same PAO.EnergyShift value) and obtained $E_{coh}(\alpha)$=3.15 eV/atom, $\Delta E_{coh}$ = 0.018 eV/atom with $U$=+1.25 eV and $E_{coh}(\alpha)$=3.17 eV/atom, $\Delta E_{coh}$ = 0.022 eV/atom with $U$=+1.0 eV. With $U$=0, $\Delta E_{coh}$ was 0.039 eV/atom with the same TZP basis.

The effect of a +$U$ value applied to $s$ states on the band structure can be seen from the PDOS (partial density of states) plot in Fig. 4, where we plot the density of $s$-like and $p$-like state. The PDOS are plotted up to the Fermi level, and $E$=0 refers to the average electrostatic potential. As expected, the +$U$ value lowers the energy of the $s$ band and affects little the $p$ band. The population averaged energy of the $s$ band (i.e. $<s> = \int_{-\infty}^{E_f} E \times PDOS(E) dE$) changes from -8.85 eV to -9.38 eV in α Sn and from -10.08 eV to -10.66 eV in β Sn when going from $U$=0 to $U(s)$=+1.75 eV. Simultaneously, the $p$-$s$ gap increases from 0.26 to 1.06 eV in α Sn and from 2.49 to 2.99 eV in β Sn. The effect on the Fermi level is minimal: within 0.05 eV for both phases when $U$ is increased from 0 to 1.75 eV; the changes in the average electrostatic potential are negligible when applying a $U$ value of this magnitude.

We have therefore shown that (i) the application of +$U$ affects similarly the band structure of tin atom and dimers and of bulk Sn; (ii) in tin atoms and dimers, the $s$ band energy is destabilized more than the *p* band energy by DFT vs CCSD(T); (iii) alpha tin has a smaller fraction of *s* electrons than beta tin. Together (i)-(iii) allow us to conclude that (i) DFT with the PBE functional over-stabilizes alpha over beta phase of tin by the mechanism of destabilizing *s* band more than the *p* band; (ii) this can be fixed by applying a positive *U* value to *s*-like states which permits obtaining accurate cohesive energies of both phases with the same computational setup.

We have confirmed this result by performing calculations in VASP using a different type of basis set and of pseudopotential (i.e. plane waves and PAW, respectively). The results are also summarized in Table 1. Similarly to SIESTA calculations, the *s* population is higher in the beta phase by about 0.2 *e*/atom. The application of *U*=+1 eV on *s* levels does lead to the shift down in energy of the *s*–like band while inducing no significant shift in the *p*-like band (see Fig. A1 in the Appendix) and results in the change in $\Delta E_{coh}$ from 0.039 eV/atom to 0.023 eV/atom. This simultaneously improves the cohesive energy of α Sn from 3.20 eV/atom to 3.16 eV/atom. That is, also in VASP we are able to choose a +*U*(*s*) value that results in correct cohesive energies of both α and β Sn. Similar values of +*U*(*s*) achieve $\Delta E_{coh}$ of about 0.02 eV/atom in both VASP and SIESTA. That optimal values of +*U* a little different in VASP and SIESTA is not surprising given the somewhat differing definitions of *s* and *p* character with plane waves and with atom-centred bases and otherwise significant differences in the computational approaches and implementation. What is important is the common to these setups mechanism of destabilization of *s* levels vs *p* levels which leads to an overestimation of $\Delta E_{coh}$ and which can be fixed by applying a +*U* value on *s*-like levels which allows reproducing accurate cohesive energies of both alpha and beta tin with one and the same computational scheme.

## Conclusions

In the *ab initio* modelling of tin based materials, specifically doped tin and alloys, it is necessary to be able to correctly model the energetics of the two low temperature and low pressure phases, namely, alpha and beta Sn, with the same computational setup. The small difference in the cohesive energies of these phases (about $\Delta E_{coh}$=0.02 eV/atom) and the resulting proximity of phase transition temperature to the room temperature make such modelling difficult. Specifically, the most practical and by far the most widely used approximation – DFT with the PBE functional – overestimates the difference in cohesive energies. We have shown that values of $\Delta E_{coh}$≈0.04 eV/atom are obtained with different types of basis sets and with different models of core electrons (all-electron or pseudopotentials of different types), which also leads to a significant overestimation of the α–β transition temperature.

To identify the reason for this overestimation, we have analysed for the first time possible contributions from anharmonicity to the vibrational part of the free energy as well as errors in the electronic structure. We have shown that anharmonicity does not have a significant effect on $T_{\alpha\beta}$. By comparing DFT and CCSD(T) calculations on small model systems, we have shown that DFT with the PBE functional overestimates the energy of the *s* band to a larger extent than that of the *p* band. As *s* population is higher in beta Sn than in alpha Sn, this leads to overstabilization of α vs β Sn in DFT calculations. We have shown that it is possible to correct for this by applying a positive *U* value on *s* levels, whereby correct cohesive energies of both α and β Sn can be obtained with the same computational scheme.

The cohesive energies of α-Sn and β-Sn is therefore a highly practically relevant example of the need for developing of computationally efficient exchange correlation functionals that perform well for different phases of the same material with qualitatively different electronic structure (such as metalling and non-metallic phases).

## Acknowledgements

This work was supported by the Ministry of Education of Singapore via an AcRF grant (R-265-000-494-112). The authors are grateful to the University of Oslo for hospitality, financial support and access to the Abel cluster, owned by the University of Oslo and the Norwegian Metacenter for Computational Science (NOTUR), and operated by the Department for Research Computing at USIT, the University of Oslo IT-department. We thank Prof. Clas Persson from the University of Oslo for hospitality and discussions. Dr. Oleksandr Malyi is thanked for useful discussions and for help with some of the calculations.

## Appendix

**Details of the effect of *U*(*s*) on the electronic structure of a Sn atom and a Sn dimer computed in VASP**

The effect of lowering the *s* band energy and of increasing the *p*-*s* gap was confirmed in Sn atom and dimer calculations in VASP. For example, applying *U*(*s*) = 1.5 eV reduces the destabilization of the *s* band of a Sn atom vs. CCSD(T) from 3.1 eV to 2.8 eV while the destabilization of the *p* states remains at 3.0 eV; the *s* band destabilization in the dimer is reduced from 4.0 eV to 3.7 eV and the degree of underestimation of the *p*-*s* gap from 1.6 eV to 1.2 eV, while the *p* band destabilization remained at 2.4 eV.

**Details of band structure of α and β tin computed in VASP**

The VASP calculations assign populations for alpha/beta Sn of 1.84/2.04 for the *s* states (and thus 2.16/1.96 for the *p* states). Similarly to SIESTA calculations (and to what expected), the metallic beta Sn phase allocates more electrons to the *s* bands than the semiconducting alpha Sn phase does (by 0.2 *e*/atom). VASP calculations show that applying a Hubbard correction of magnitude +1.0 to *s* states improves notably the energy cohesive difference between the α and β phases. Analysis of the calculations shows that by applying *U*(*s*)=1.0, the population averaged energy for *s* states is decreased by 0.09 eV for alpha Sn (from -7.11 to -7.20 eV), and by 0.14 eV for beta Sn (from -6.61 to -6.75 eV), while that for *p* states is slightly increased: by 0.04 eV for alpha Sn (from -2.55 to -2.51 eV) and by 0.03 eV for beta Sn (from -2.54 to -2.51 eV). Here, the energy values refers to $E_f$ = 0 eV. A positive Hubbard $U(s)$ correction therefore stabilizes (very slightly destabilizes) the $s$ ($p$) bands for alpha and beta Sn.

The plots of the projected density of states with ($U$=1.0) and without ($U$=0) a Hubbard correction to $s$ levels are given in Figure A1 for the two phases of Sn. The PDOS is plotted by setting the Fermi energy to zero. Similar to SIESTA calculations (see main text), the application of $U$ only changes the vacuum and the Fermi levels by less than 0.05 eV when going from $U$=0 to $U$=1.5 eV. The shift of the $s$ bands lower with respect to $p$ bands for $U$=1.0 versus $U$=0 is visible for both alpha and beta Sn.

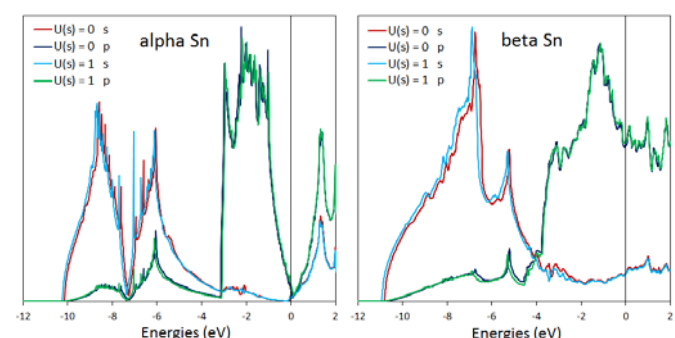

**Fig. A1** Projected density of states for alpha (left) and beta (right) Sn with (U=1) and without (U=0) Hubbard correction to $s$-like levels.

## Notes and references

1　S. Groves and W. Paul, *Phys. Rev. Lett.*, 1963, **11**, 194.
2　Y. Wang, M. Wu, Z. Jiao, and J. Y. Lee, *Chem. Mater.*, 2009, **21**, 3210.
3　L. D. Ellis, T. D. Hatchard, and M. N. Obrovac, *J. Electrochem. Soc.*, 2012, **159**, A1801.
4　L. D. Ellis, P. P. Ferguson, and M. N. Obrovac, *J. Electrochem. Soc.*, 2013, **160**, A869.
5　N. Singh, T. S. Arthur, C. Ling, M. Matsui, and F. Mizuno, *Chem. Commun.*, 2013, **49**, 149.
6　M. N. Obrovac and V. L. Chevrier, *Chem. Rev.*, 2014, **114**, 11444.
7　B. Wang, B. Luo, X. Li, and L. Zhi, *Mater. Today*, 2012, **15**, 544.
8　K. Hirai, T. Ichitsubo, T. Uda, A. Miyazaki, S. Yagi, and E. Matsubara, *Acta Mater.*, 2008, **56**, 1539.
9　H. S. Im, Y. J. Cho, Y. R. Lim, C. S. Jung, D. M. Jang, J. Park, F. Shojaei, and H. S. Kang, *ACS Nano*, 2013, **7**, 11103.
10　D. C. Langreth and M. J. Mehl, *Phys. Rev. B*, 1983, **28**, 1809.
11　J. P. Perdew, K. Burke, and M. Ernzerhof, *Phys. Rev. Lett.*, 1996, **77**, 3865.
12　P. Haas, F. Tran, and P. Blaha, *Phys. Rev. B*, 2009, **79**, 085104.
13　P. Janthon, S. Luo, S. M. Kozlov, F. Viñes, J. Limtrakul, D. G. Truhlar, and F. Illas, *J. Chem. Theory Comput.*, 2014, **10**, 3832.
14　A. D. Becke, *J. Chem. Phys.*, 1993, **98**, 1372.
15　J. P. Perdew, M. Ernzerhof, and K. Burke, *J. Chem. Phys.*, 1996, **105**, 9982.
16　P. Deák, B. Aradi, and T. Frauenheim, *Phys. Rev. B*, 2011, **83**, 155207.
17　A. Jain, S.P. Ong, G. Hautier, W. Chen, W.D. Richards, S. Dacek, S. Cholia, D. Gunter, D. Skinner, G. Ceder, K.A. Persson, *APL Mat.*, 2013, **1**, 011002.
18　P. Pavone, S. Baroni, and S. de Gironcoli, *Phys. Rev. B*, 1998, **57**, 10421.
19　A. D. Becke, *J. Chem. Phys.*, 1993, **98**, 5648.
20　A. I. Liechtenstein, V. I. Anisimov, and J. Zaanen, *Phys. Rev. B*, 1995, **52**, R5467.
21　G. Kresse and J. Furthmüller, *Comput. Mater. Sci.*, 1996, **6**, 15.
22　http://elk.sourceforge.net
23　J.M. Soler, E. Artacho, J.D. Gale, A. Garcia, J. Junquera, P. Ordejon, D. Sanchez-Portal, *J. Phys. Condens. Matter* **14**, 2745 (2002).
24　V. Blum, R. Gehrke, F. Hanke, P. Havu, V. Havu, X. Ren, K. Reuter, and M. Scheffler, *Comput. Phys. Commun.*, 2009, **180**, 2175.
25　G. Kresse and D. Joubert, *Phys. Rev. B*, 1999, **59**, 1758.
26　K. Lejaeghere, V. Van Speybroeck, G. Van Oost, and S. Cottenier, *Crit. Rev. Solid State Mater. Sci.*, 2013, **39**, 1.
27　N. Troullier and J. L. Martins, *Phys. Rev. B*, 1991, **43**, 1993.
28　C. Kittel, Introduction to Solid State Physics, 8th ed. (Wiley, Hoboken, NJ, 2005). p.50
29　M. Wierzbowska, D. Sánchez-Portal, and S. Sanvito, *Phys. Rev. B*, 2004, **70**, 235209.
30　S. L. Dudarev, G. A. Botton, S. Y. Savrasov, C. J. Humphreys and A. P. Sutton, *Phys. Rev. B*, 1998, **57**, 1505.
31　E. van Lenthe, A. Ehlers, and E.-J. Baerends, *J. Chem. Phys.*, 1999, **110**, 8943.
32　J. A. Pople, M. Head-Gordon, and K. Raghavachari, *J. Chem. Phys.*, 1987, **87**, 5968.
33　M. J. Frisch et al., Gaussian 09, Gaussian, Inc., Wallingford CT, 2009.
34　P. J. Hay and W. R. Wadt, *J. Chem. Phys.*, 1985, **82**, 299.
35　W. R. Wadt and P. J. Hay, *J. Chem. Phys.*, 1985, **82**, 284.
36　B. Metz, H. Stoll, and M. Dolg, *J. Chem. Phys.*, 2000, **113**, 2563.
37　J. Ihm and M. L. Cohen, *Phys. Rev. B*, 1981, **23**, 1576.
38　P. Ehrhart, Landolt-Börnstein - Group III Condensed Matter, Vol. 25.
39　S. Manzhos, X. Wang, R. Dawes, and T. Carrington, *J. Phys. Chem. A*, 2006, **110**, 5295.